\documentclass[11pt,twoside]{article}

\usepackage{asp2006}
\usepackage{epsf}
\usepackage{psfig}
\usepackage{lscape}

\def\omh{\Omega_{\rm m} h}
\def\omb{\Omega_{\rm b}}
\def\om{\Omega_{\rm m}}

\def \bj {b_{\rm J} }

\begin{document}
\title{The Galaxy Power Spectrum: 2dFGRS-SDSS tension?}   
\author{Shaun Cole\altaffilmark{1}, Ariel G. S\'anchez\altaffilmark{1,2}, Stephen Wilkins\altaffilmark{1}}   
\altaffiltext{1}{University of Durham, South Road, Durham, DH1 3LE}
\altaffiltext{2}{Grupo de Investigaciones en Astronom\'ia Te\'orica y
  Experimetal (IATE), OAC, UNC, Argentina}    

\begin{abstract} 
Published galaxy power spectra from the 2dFGRS and SDSS
are not in good agreement. We revisit this issue by
analyzing both the 2dFGRS and SDSS DR5 catalogues
using essentially identical technqiues. We confirm
that the 2dFGRS exhibits relatively more large scale power
than the SDSS, or, equivalently, SDSS has more small scale 
power. We demonstrate that this difference is due 
the $r$-band selected SDSS catalogue being dominated by
more strongly clustered red galaxies,
due to these galaxies having a stronger
scale dependent bias. The power spectra of galaxies
of the same rest frame colours from the two surveys
match well. It is therefore important to accurately model
scale dependent bias to get accurate estimates of cosmological
parameters from these power spectra.
\end{abstract}



\section{Introduction}   

       Measurements of large scale galaxy clustering place important
constraints on cosmological parameters that complement those from
the analysis of fluctuations in the cosmic microwave background (CMB).
Measurments of the galaxy power spectrum from SDSS \citep{tegmark04}
and 2dFGRS \citep{cole05} constrain the parameter combinations
$\omh$ and $\omb/\om$. The constraint on $\omh$ is particularly
important as, for instance, it breaks a degeneracy in the CMB data  
and allows accurate determination of $\om$. Hence to be sure
that systematic errors are not biasing the parameters it is very
useful to have independent estimates from different surveys.
However when one looks at the constraints coming from the published
2dFGRS and SDSS analysis one finds a tension. In figure~16a of
\citet{cole05}, which compares the published estimates of the galaxy
power spectra, there is evidence of more large scale power in
the 2dFGRS than in SDSS. This folds through and results in the
headline value of $\omh =0.168\pm 0.016$ from 2dFGRS \citep{cole05}
being lower than that of \citet{tegmark04} SDSS, $\omh = 0.213\pm 0.023$.
Here one should be cautious as different priors have been assumed,
but in the analysis of \citet{sanchez06}, which treats each data set on
an equal footing and separatley combines each
with CMB data, 
one sees in their figure~18 that the SDSS prefers a substantially
higher value of $\om$ than does the 2dFGRS. In fact, while the
2dFGRS estimate is in good agreement with that from the CMB data alone
(it significantly tightens the constraint without shifting the
best fitting value) the SDSS data pull $\om$  to higher
values than preferred by the CMB.

Here we seek to investigate whether these differences
are as a result of: a)~larger than expected cosmic variance, 
b)~systematics due to differences in the analysis technique (Cole et al.
use simple Fourier methods in redshift space, while Tegmark et al. use
the apparatus of KL decomposition and work in real space), 
c)~systematics due to problems with galaxy catalogues or d) intrinsic
differences in the underlying galaxy clustering.  In order to directly
compare the 2dFGRS and SDSS, we analyse each dataset using essentially
identical methods which we outline in Section~\ref{sec:method}  In
Section~\ref{sec:results} we briefly look at the region of overlap
between the surveys and then compare the resulting power spectra from
the full catalogues and interpret the differences.  We conclude in
Section~\ref{sec:conc}

\section{Methods}
\label{sec:method}  

 \subsection{2dFGRS analysis}   

  Our method of estimating the galaxy power spectrum, determining
statistical errors and fitting models
is essentially identical to that set out
in \citet{cole05}, but with three minor changes. In brief:
\begin{itemize}
      \item{} We use masks, whose construction is described in 
\citet{norberg02}, to describe the angular variation of the survey magnitude
limit, redshift completeness and magnitude dependence of the redshift
completeness.
      \item{} Random catalogues are generated by sampling from the
luminosity function and viewing through the masks. To generate
random catalogues corresponding to red/blue subsets, the luminosity
function of only the red/blue galaxies is used.
      \item{} Close pair incompleteness due to ``fibre collisions''
is dealt with by redistributing the weights of missed galaxies to their
10 nearest neighbours on the sky.
      \item{} The power spectrum is estimated using a simple cubic 
FFT method with the optimal weighting scheme of \citet[PVP]{PVP} and then
spherically averaged in redshift space. The assumed linear 
empirical bias factors that are used in this weighting scheme are
\begin{eqnarray}
b_{\rm blue} &=0.9\, (0.85+0.15\, L/L_*)\qquad    {\rm for\ rest\
  frame\ } b_{\rm J} -r_{\rm F} <1.07 \cr
b_{\rm red} &= 1.3\, (0.85+0.15\, L/L_*) \qquad     {\rm for\ rest\
  frame\ } b_{\rm
  J} -r_{\rm F} >1.07 .
\label{eqn:bias}
\end{eqnarray}
      \item{} The covariance matrix describing the errors
on the power spectrum measurements and their correlations is 
estimated using mock catalogues which are constructed from the 
random catalogues by generating a log-normal density field with
a specified power spectrum
and using it to modulate the selection of galaxies from the
random catalogue. 
Thus, by construction, these catalogues have a power spectrum very
close to the best fitting model and have luminosity and colour
dependent clustering consistent with the bias factors of
equation~(\ref{eqn:bias}).
      \item{} The survey window function is determined directly from
the random catalogue. When fitting models the theoretical model power
spectra are convolved with the survey window function.
\end{itemize}

\begin{figure}[h]
\centerline{\psfig{figure=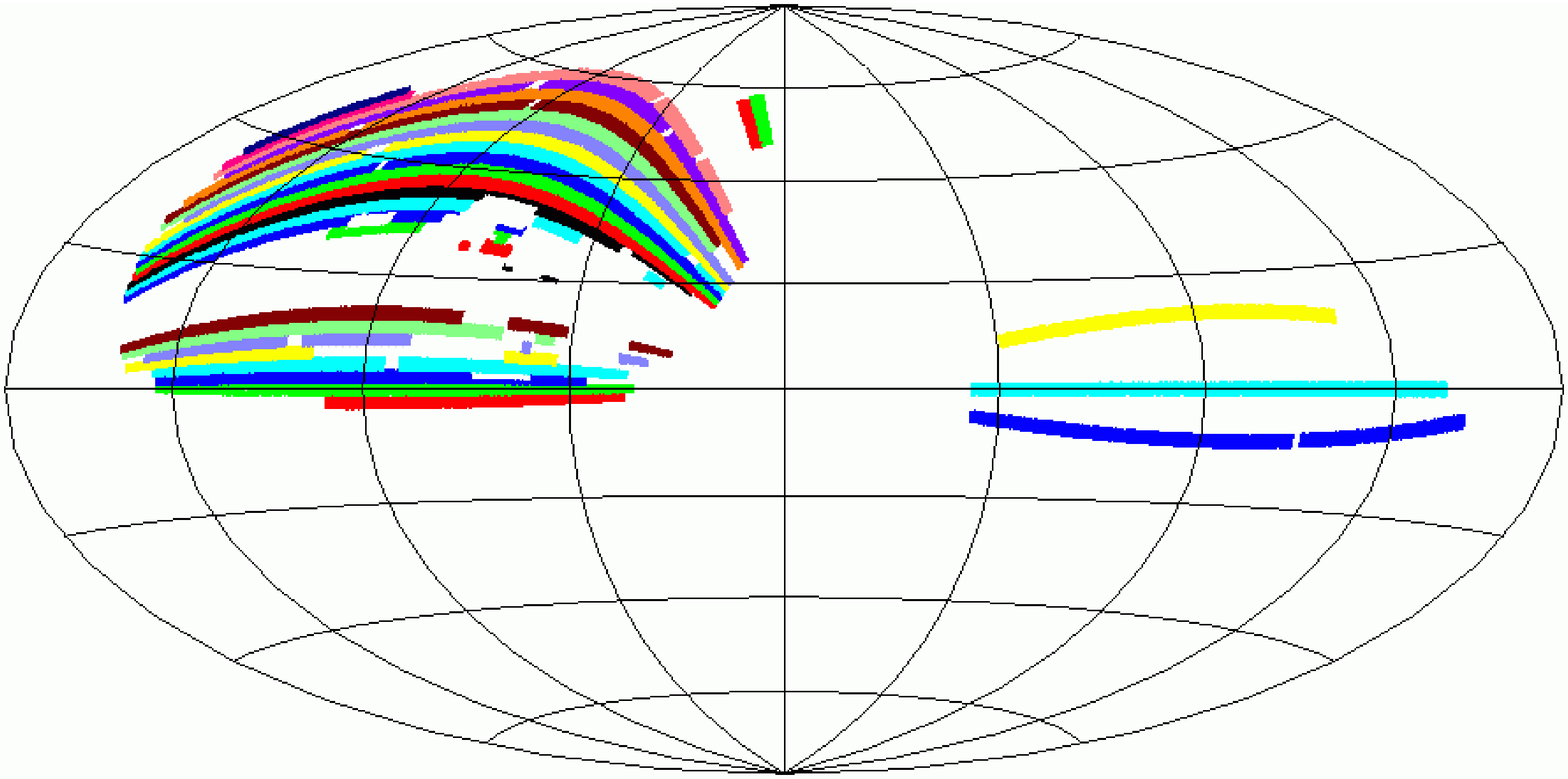,width=11cm}}
\centerline{\psfig{figure=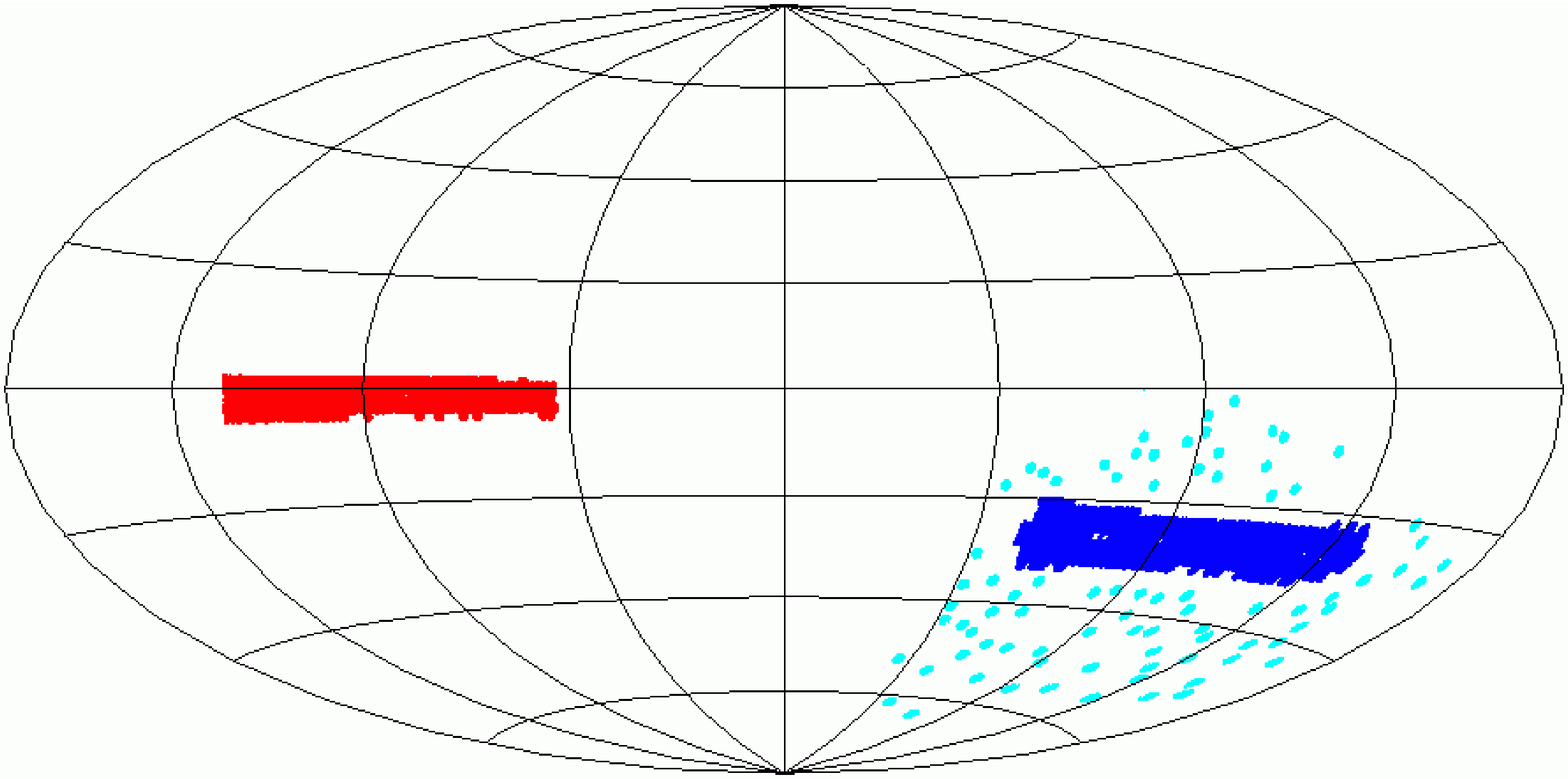,width=11cm}}
\caption{Equal area all sky Aitoff projections of all the
galaxies accepted by our SDSS masks (upper) and
2dFGRS mask (lower) and used in our analysis.}
\label{fig:mask}
\end{figure}

The three minor changes we have made are:
\begin{itemize} 
    \item{} We have changed the binning scheme so that now $P(k)$
is estimated in bins uniformly spaced in $\log_{10} k$, rather
than the linearly space bins with different bin widths in different
ranges of $k$ that were used in \citet{cole05}.
    \item{} We used new sets of log-normal catalogues in which 
the modulation of the density field used for galaxies with different 
bias factors is linear rather than the slightly more complicated 
scheme that was employed by \citet{cole05}.
    \item{} Power spectra fits are done using Cosmo-MC \citep{lewis}
as in \citet{sanchez06} which
uses CAMB to estimate of $P(k)$ rather than the \citet{EH98} fitting formula.
(We have found the use of the approximate  \citet{EH98} 
fitting formula causes a small shift in $\omh$.)
\end{itemize} 

 \subsection{SDSS analysis}   

In most respects our analysis of the SDSS is identical to that of
the 2dFGRS. The only differences are a simpler way of generating 
the survey masks and of populating the corresponding random catalogue.

The sky coverage mask we have adopted for the SDSS-DR5 data
is shown in Fig.~\ref{fig:mask} and compared with corresponding
2dFGRS mask. We constructed this mask by simply noting the 
angular coverage of each of the stripes from which the SDSS
survey is built and by removing a few small regions with poor
coverage. Most of the SDSS survey goes to a uniform magnitude limit of
$r=17.77$, but a sub-area, which is easily identified using the
target selection date, has a variety of different magnitude limits.
In this sub-area we simply imposed a fixed magnitude limit of
$r=17.5$ and discarded all galaxies fainter than this limit.
The number of galaxies with redshifts that are retained
by the mask and magnitude limits and used in our analysis is $443\, 424$ .
The mask is cruder than
the more sophisticated ones employed by \citet{tegmark04} and for
the 2dFGRS as it ignores the smaller scale variation in the redshift
completeness. However in the case of the 2dFGRS, where the
incompleteness variation is much larger, we showed that
provided this incompleteness is accounted for using our method of
redistributing of the weights of galaxies without redshifts to
neighbours with redshifts the resulting power spectrum estimates
are very accurate 
\citep[see figure~17g of ][]{cole05}.

\begin{figure}[h]
\psfig{figure=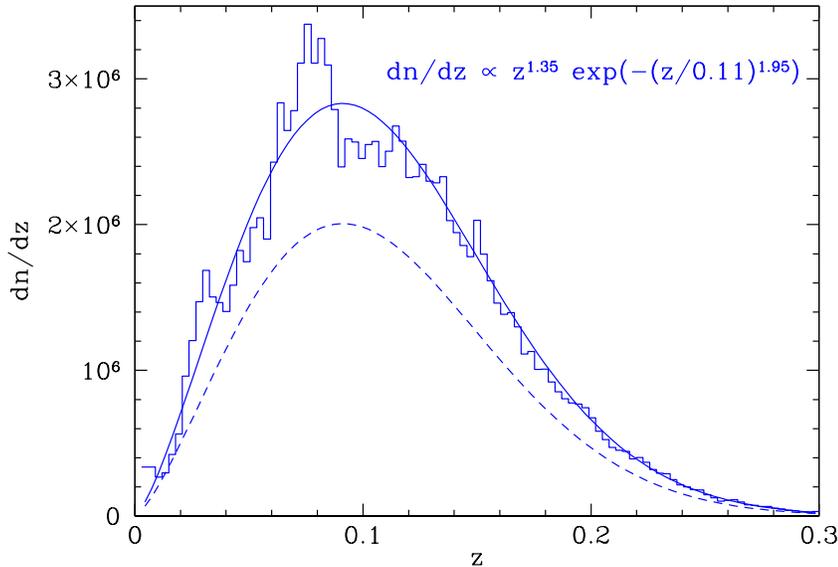,width=12cm,angle=270
,bbllx=37bp,bblly=0bp,bburx=540bp,bbury=773bp}
\caption{The redshift distribution of the SDSS $r<17.77$ sample.
The solid line shows the analytic fit
used to generate the corresponding random catalogue. The dashed
line is this same fit scaled down in amplitude to be always below
the redshift histogram.
}
\label{fig:dndz}
\end{figure}

As most of the catalogue goes uniformly to the deeper
magnitude limit a simple method can be used to  construct the
random catalogue. Fig.~\ref{fig:dndz} shows the redshift distribution
of this sample together with an analytic fit  that smoothes away
the effect of large scale clustering. Our procedure is:
\begin{enumerate}
\item{}Select a random direction on the sky.
\item{}Choose at random a genuine galaxy from the region of
  the survey that goes to $r=17.77$ .
\item{}Keep the galaxy with a probability proportional to the ratio
 of the height of redshift histogram to that of the scaled fit 
 shown as the dashed line in Fig.~\ref{fig:dndz}
 at the redshift of the selected galaxy .
\item{}Keep or discard the galaxy based on the sky coverage and
magnitude limit of the mask. (Note, for galaxies that fall where
the  magnitude limit is only $17.5$, the fainter
galaxies will be discarded and the redshift distribution of the
retained galaxies will be appropriately shallower than that of
Fig.~\ref{fig:dndz}.)

\end{enumerate}
These steps are done repeatedly until a random catalogue
containing $100$ times more galaxies than the genuine catalogue
is built up. Selecting from the genuine catalogue in this way
means we automatically have apparent magnitudes and colours for
all the galaxies in the random catalogue and so can select sub-samples
from it and weight its galaxies in just the same way as the genuine
catalogue.

To utilize the \citet{PVP} optimum weighting we need to 
determine bias factors for the galaxies in 
the genuine, random and mock catalogues.
We do this  by first converting the SDSS
magnitudes to the 2dFGRS $b_J$ and $r_F$ bands using 
\begin{eqnarray}
b_{\rm J} &=& g + 0.15 +0.13\, (g-r) \cr
r_{\rm F} &=& r -0.13
\label{eqn:col}
\end{eqnarray}
and the simple colour dependent $k$-corrections that were
used for the 2dFGRS data \citep[see section~3 of][]{cole05}.
Then we are able to define the bias factors using
equation~(\ref{eqn:bias}) just as for the 2dFGRS data.

\section{Results}
\label{sec:results}

Our main focus is the comparison of the power spectra of the
two surveys, but we first directly compare the two surveys
in the region of their overlap to get a feel for the different
selections used and the level of incompleteness.

\subsection{Survey overlap}   

\begin{figure}[t!]
\psfig{figure=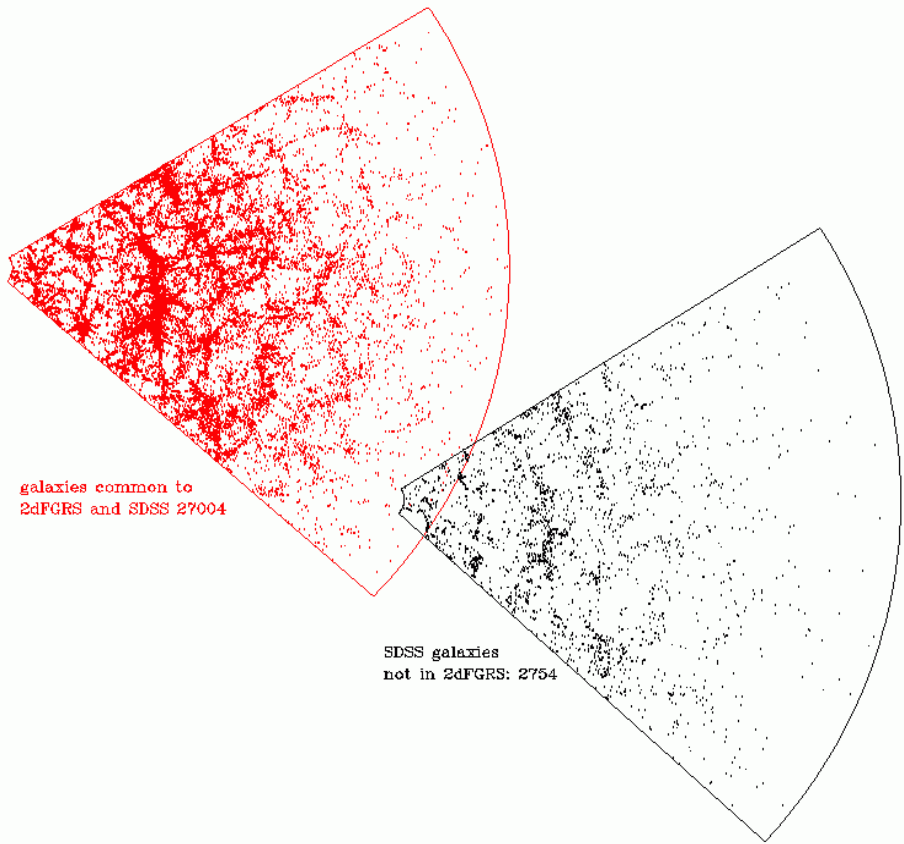,width=12cm,bbllx=-65bp,bblly=65bp,bburx=670bp,bbury=760bp}
\caption{Cone plots showing RA and redshift for galaxies 
in the region of sky where the 2dFGRS and SDSS surveys overlap.
There are $53\, 382$  within this area that are in both
surveys and in the upper panel we plot the sub-sample of  $27\, 004$
that (estimated from SDSS photometry) have $b_{\rm J}<19$
and redshift $z>0.01$. 
The lower panel shows the $2754$ galaxies that are in SDSS and pass
the same magnitude and redshift cuts, 
but are missing in the 2dFGRS catalogue.
}
\label{fig:cones} 
\end{figure}

In the northern galactic hemisphere there is a contiguous area of
overlap between the two surveys, which runs for $74$~degrees of RA
and is for the most part $5.2$~degrees wide in declination. 
If we select from the SDSS photometric catalogue all galaxies
brighter than $\bj=20$ (we do not apply the $r \approx 17.77$
magnitude limit of the SDSS main galaxy survey, but we do apply
all the other star-galaxy classification criteria used in that sample
\citep[see][]{strauss02}), then in this area there are $53\, 382$ galaxies that are in 
both catalogues. We find the fraction of SDSS galaxies which are
also in the 2dFGRS to be constant at $89$\% as faint as $\bj \approx
18.9$. Fainter than this, SDSS galaxies are missing from the 2dFGRS
sample simply due to the (variable) magnitude limit of the 2dFGRS
survey and its $0.15$~magnitude random photometric errors.
This finding is in perfect accord with the estimates 
made with the SDSS EDR \citep{stoughton} in \citet{norberg02}.
This $11$\% incompleteness has been investigated by
\citep{cross} as well as \citet{norberg02} and has been shown to be
predominately due to incorrect star-galaxy classification.
The star-galaxy classification parameters based on the
APM photometry are noisy and this level of 
incompleteness  is in line with what was expected \citep{maddox90}.

\begin{figure}[t!]
\psfig{figure=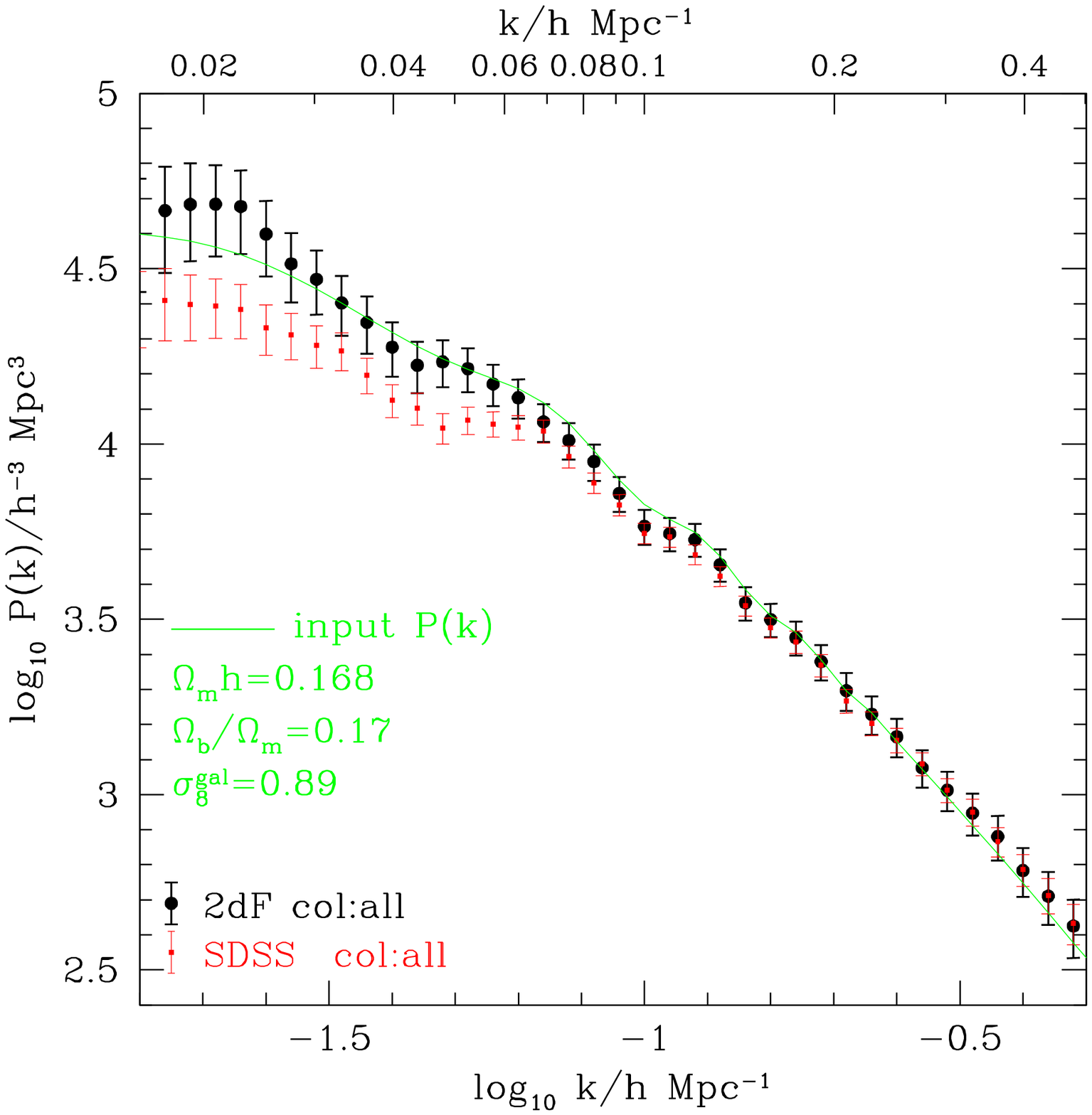,bbllx=0bp,bblly=12bp,bburx=573bp,bbury=560bp,width=13cm,rheight=1.2cm} 
\rightline{\psfig{figure=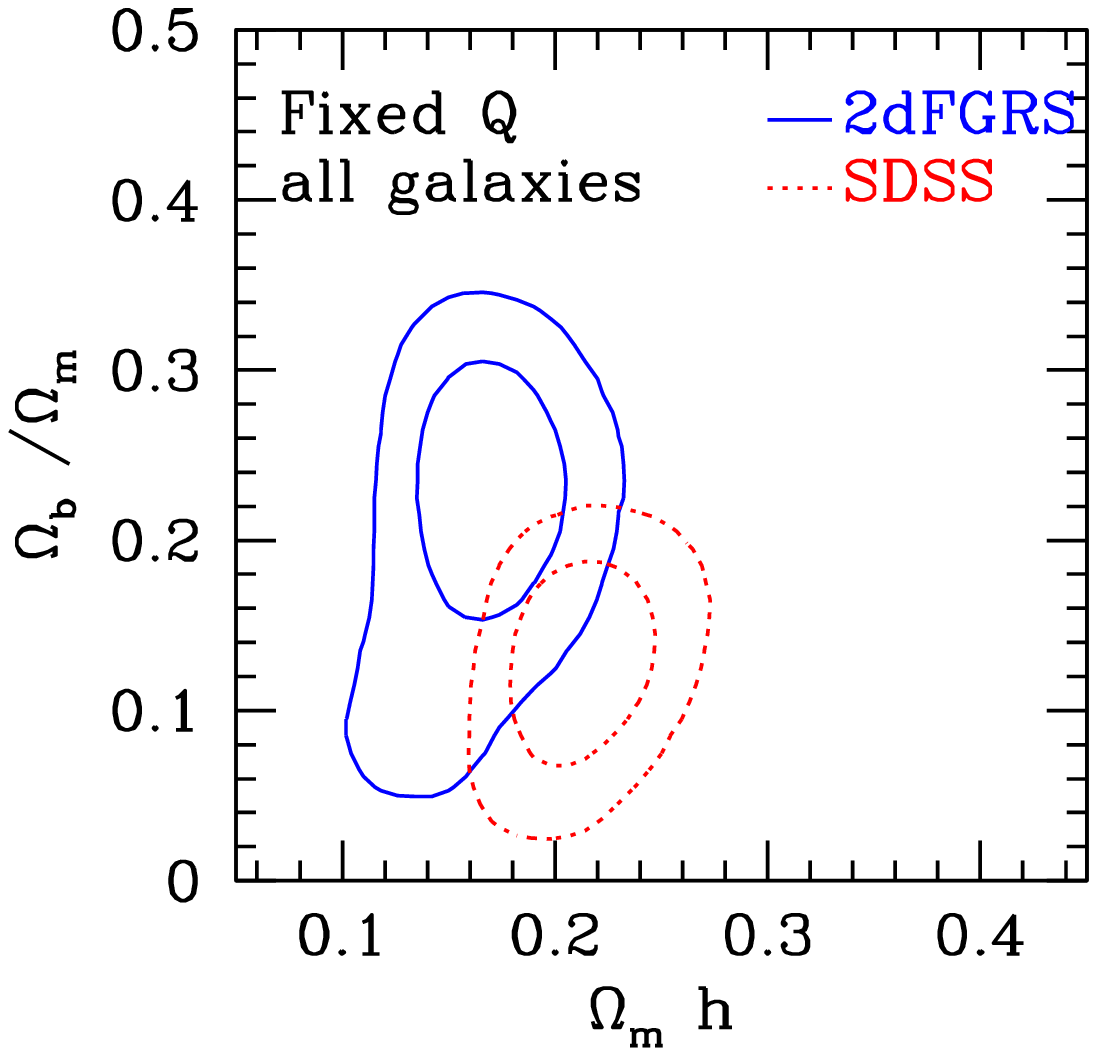,bbllx=246bp,bblly=390bp,bburx=573bp,bbury=704bp,width=5.0cm,rheight=11.0cm}
\qquad \quad}
\caption{Comparison of the power spectra estimated from
the full 2dFGRS and SDSS DR5 samples - corrected for
the effect of the window function as described in the text. 
Inset: The contours show 68\% and 95\%
joint confidence intervals for the baryon fraction
$\Omega_{\rm b}/\Omega_{\rm m}$ and $\Omega_{\rm m} h$
for fits to this SDSS and 2dFGRS data in the range
$0.02<k<0.15\ h$Mpc$^{-1}$. 
The parameter $Q$
modelling the distortion of power spectrum due to nonlinearity,
redshift space distortions and scale dependent bias was kept 
fixed at $Q=5$ in these fits. 
}
\label{fig:pk_all}
\end{figure}

The issue here is whether this incompleteness has any influence
on estimates of galaxy clustering. We can look at this
directly by plotting cone plots (Fig.~\ref{fig:cones}) of the
galaxies the two catalogues have in common and those missed
by the 2dFGRS. Here we plot only galaxies with $\bj<19$ to avoid
issues with the 2dFGRS magnitude limit.
Here we see that $91$\% of the SDSS galaxies are in the 2dFGRS. 
Comparing the two cone plots in Fig.~\ref{fig:cones}, it
appears that the galaxies missed by 2dFGRS are just a random
sparse sampling of the structure seen in the  matching sample
and so there is no evidence that the incompletness
has a spatial imprint.

\subsection{Comparison of power spectra}

\begin{figure}[ht]
\psfig{figure=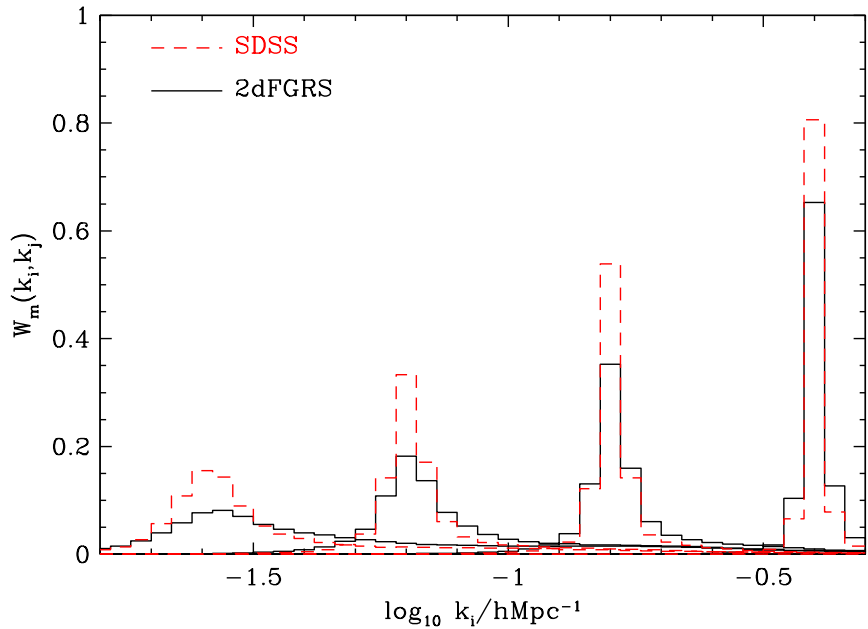,width=12cm,bbllx=0bp,bblly=363bp,bburx=537bp,bbury=762bp}
\caption{A sample of the window functions for our individual
band power estimates for both SDSS (red dashed)  
and 2dFGRS (black solid).
}
\label{fig:wins}
\end{figure}

In Fig.~\ref{fig:pk_all} we compare the 
`deconvolved' power spectra estimated
from the full 2dFGRS catalogue and full SDSS-DR5 sample using
the methods we outlined in Section~\ref{sec:method}
The power spectra we estimate are the underlying galaxy power
spectra convolved with the window function of either the
SDSS of 2dFGRS as appropriate
\begin{equation}
\hat P({\bf k}) = P({\bf k}) \otimes W^2({\bf k}) .
\end{equation}
Our random catalogues allow us to accurately estimate 
$W^2({\bf k})$ and from it determine the matrix of window
functions that describe how our spherically averaged band
power estimates are related to unconvolved power spectrum
\begin{equation}
\hat P( k_j ) = \sum_i P( k_i) W_{\rm m}(k_i,k_j) .
\end{equation}
Examples of these window functions for the SDSS and 2dFGRS
are shown in Fig.~\ref{fig:wins}.
For all our quantitative analysis we use these window functions
to convolve the model power spectra before comparing with the
data. However for the purposes of visually comparing the 
2dFGRS and SDSS power spectra we have corrected the convolved
estimates by multplying them through by the ratio a similar
model power spectrum to its convolved counterpart. This
'deconvolution' is accurate provided the power spectra are smooth.

\begin{figure}[t!]
\psfig{figure=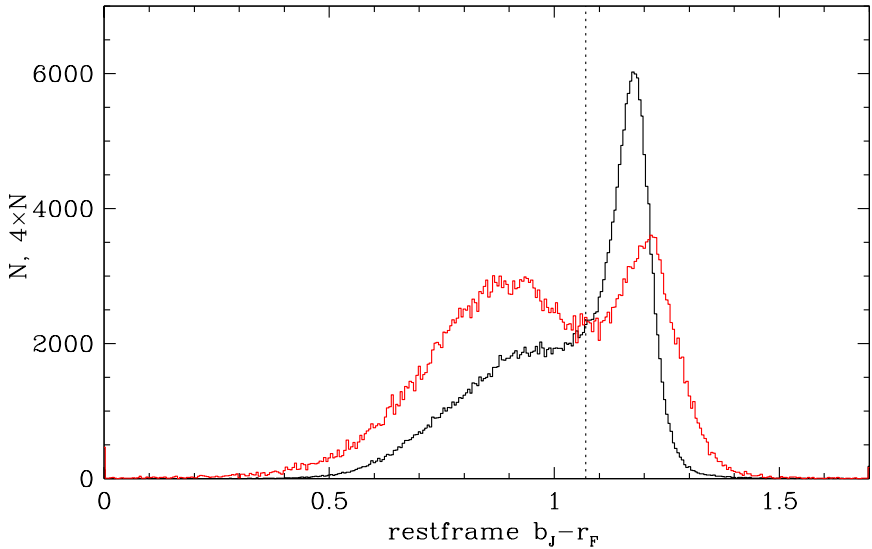,width=12cm,bbllx=0bp,bblly=55bp,bburx=537bp,bbury=393bp}
\caption{Histograms comparing the distribution of rest frame
$b_{\rm J} - r_{\rm F}$ colours in the SDSS (black) and 2dFGRS (red) 
catalogues. Note the different units on the $y$-axis.}
\label{fig:colhist}
\end{figure}

In Fig.~\ref{fig:pk_all} we note that the SDSS and 2dFGRS galaxy
power spectra agree well for $k > 0.07$~h~Mpc$^{-1}$. The good
agreement in amplitude at this wavenumber is due to the  bias
dependent weights used in the PVP estimator which have sucessfully
modelled the difference in the clustering strength of the red selected
SDSS galaxies and blue selected 2dFGRS galaxies. This is by design
as the bias factors were normalized empirically by the 2dFGRS red
and blue samples at this scale \citep[see figure~15 of][]{cole05}.
In contrast, on larger scales we see evidence for significantly
more large scale power in the 2dFGRS than in SDSS. 
Fitting power spectra of the form
\begin{equation}
P(k) = P_{\rm linear}(k,\Omega_{\rm m},h,\Omega_{\rm b}) 
\frac{1+ Q k^2}{1 + A k}
\label{eqn:nonlin}
\end{equation}
to these data yields the parameter constraints on
$\Omega_{\rm m}h$ and $\Omega_{\rm b}/\Omega_{\rm m}$
shown in inset panel of Fig.~\ref{fig:pk_all}. Here we have kept 
the parameters $A$ and $Q$, which model non-inear distortions
of the power spectrum, fixed at fiducial values of
$A=1.4$ and $Q=5$ such that the difference in shape
of the fitted power spectra are completely characterized by
the parameter combinations $\Omega_{\rm m}h$ and 
$\Omega_{\rm b}/\Omega_{\rm m}$.
We note that the 2dFGRS and SDSS best fit values lie
outside each others 95\% confidence contours and that
the SDSS parameter estimates are completely in accord with
those from \citet{tegmark04}, but with much
tighter bounds due to the larger SDSS dataset used here.
 Thus the first thing to note is that the
difference between the SDSS and 2dFGRS results that was
noted in the introduction and which motivated this analysis
is certainly significant and not an artifact of differing
analysis techniques.

\begin{figure}[t!]
\psfig{figure=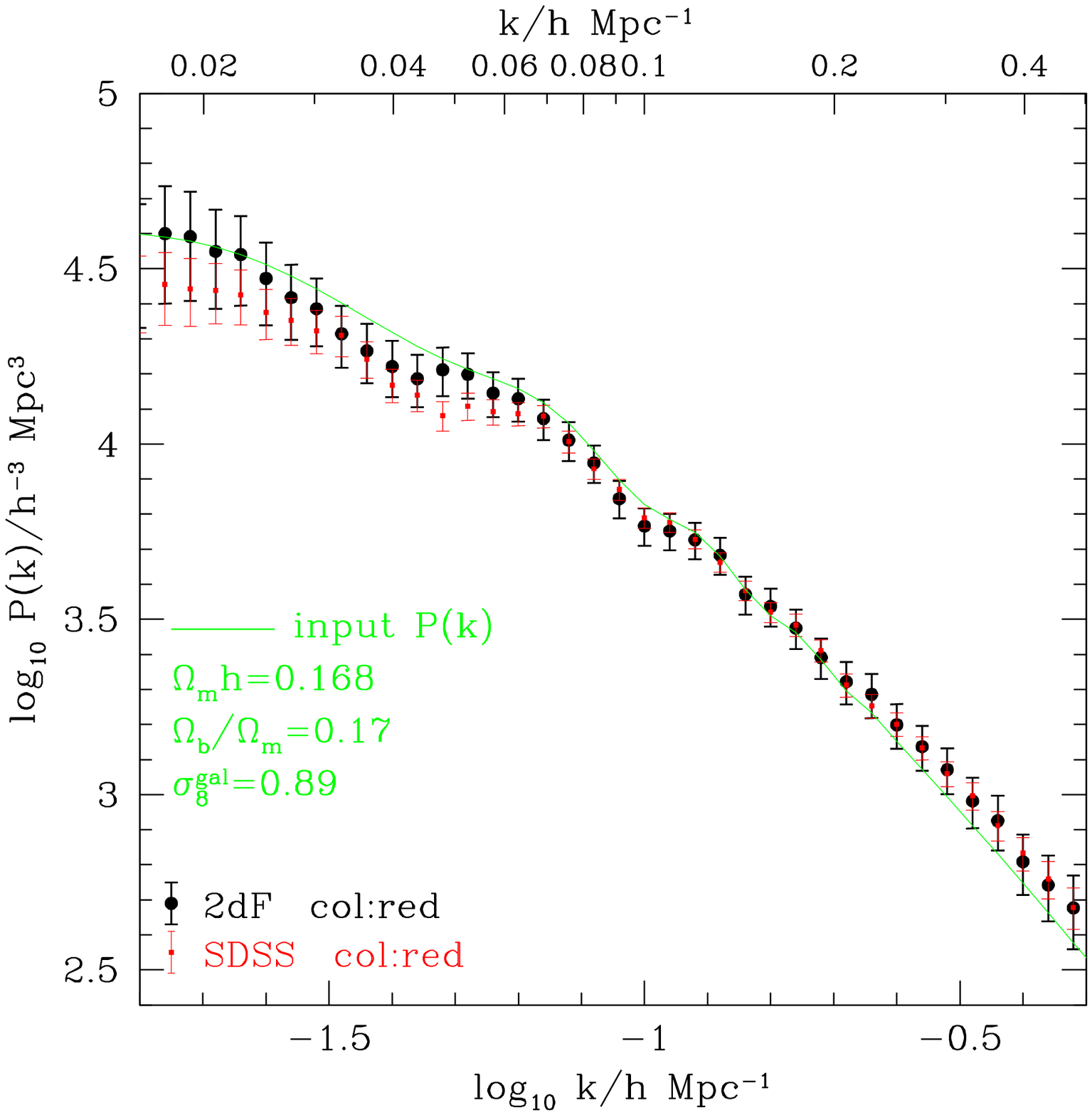,bbllx=0bp,bblly=12bp,bburx=573bp,bbury=560bp,width=13cm,rheight=1.2cm} 
\rightline{\psfig{figure=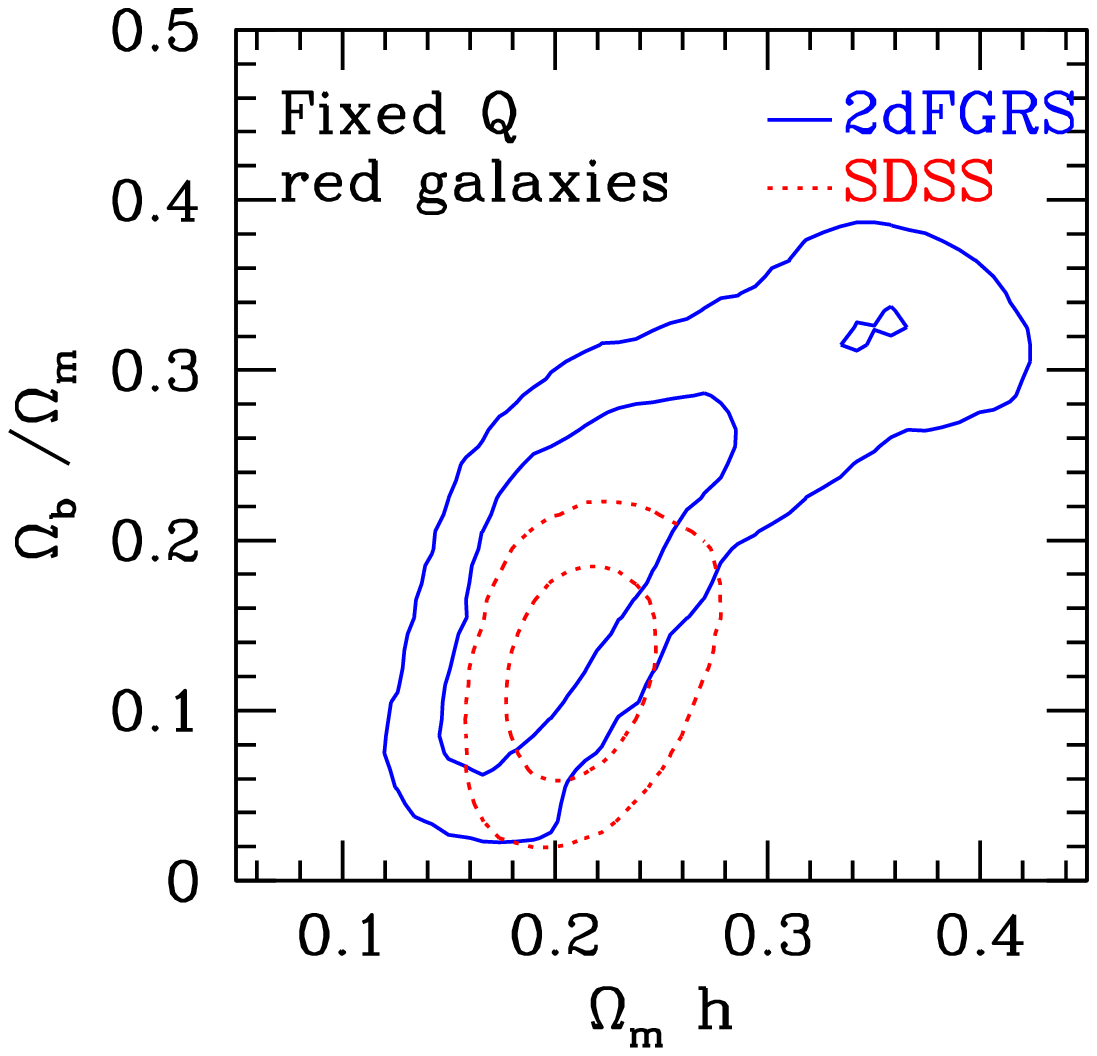,bbllx=246bp,bblly=390bp,bburx=573bp,bbury=704bp,width=5.0cm,rheight=11.0cm}
\qquad \quad}
\caption{\label{fig:pk_red}
Comparison of the 2dFGRS and SDSS-DR5 power spectra 
from red subsamples that satisfy the rest frame colour
$b_{\rm J}-r_{\rm F}>1.07$. Inset: The contours show 68\% and 95\%
joint confidence intervals for the baryon fraction
$\Omega_{\rm b}/\Omega_{\rm m}$ and $\Omega_{\rm m} h$
for fits to this  SDSS and 2dFGRS data in the range $0.02<k<0.15\ h$Mpc$^{-1}$.
Again the parameter $Q$ was kept  fixed at $Q=5$.
} 
\end{figure}

We now investigate if the discrepant shapes of the galaxy power
spectra are due to the difference in the clustering properties
of red and blue galaxies. Fig~\ref{fig:colhist} shows histograms
of rest frame  $b_{\rm J} - r_{\rm F}$ colours for both the
2dFGRS and SDSS catalogues. The SDSS magnitudes have
been converted to these bands assuming the relations given
in equation~\ref{eqn:col}. The colour distributions are clearly
bimodal with a natural dividing point at 
$b_{\rm J} - r_{\rm F}=1.07$. The 2dFGRS has roughly equal numbers
of red and blue galaxies while the SDSS, being red selected, is
naturally dominated by red galaxies.

In Fig.~\ref{fig:pk_red} we compare the galaxy spectra estimated
from just the galaxies redder than $b_{\rm J}-r_{\rm F}>1.07$ in both
samples. Comparing the SDSS $P(k)$ from just the red galaxies with 
the previous
estimate from the full SDSS catalogue reveals them to be in very
close agreement. This is to be expected as the SDSS sample is both
dominated by red galaxies and the PVP power spectrum estimator
that we employ gives them more weight than their less clustered
blue counterparts. In contrast the estimate for just the red
2dFGRS galaxies differs from that from all the 2dFGRS catalogue.
In fact it is a much closer match to the result from the SDSS data.
The only places where the two estimates are not in excellent
agreement is on the very largest scales $k<0.025$~h~Mpc$^{-1}$,
where the estimates are both noisy and highly correlated,
and also around $k\approx0.05$~h~Mpc$^{-1}$.  In fact this difference
is also due to sample variance. \citet{cole05} investigated
the effect of removing from the 2dFGRS catalogue the two
largest super clusters. Their figure~17 (panels~o and~p) shows
that this in general has a small effect, but does perturb the
power just around $k\approx0.05$~h~Mpc$^{-1}$.

Agreement, within the expected statistical uncertainty, is
confirmed in the inset panel of Fig.~\ref{fig:pk_red} where we see the best
fitting parameters for 2dFGRS lie within the SDSS 67\% confidence
contour and vice versa. Note that again we have kept the parameters
describing the nonlinear distortion fixed.

\section{Conclusions}   
\label{sec:conc}

The conclusion of this investigation is that a significant difference
exists between the shape of the galaxy power spectra measured in the
2dFGRS and SDSS surveys and that this difference is due to scale
dependent bias. If a homogenous sample of red galaxies is selected
from each survey then the resulting power spectra agree to within
the expected statistical errors. 
In contrast, when the full 2dFGRS and SDSS catalogues are analyzed the 
resulting 2dFGRS power spectrum differs in shape to that of SDSS.
If normalized on scales around $k\approx0.1$~h~Mpc$^{-1}$,
as is done automatically by the scale independent bias factors
assumed in the PVP estimator, then the 2dFGRS $P(k)$ exhibits more large
scale power than SDSS. However, if one instead normalizes on large
scales one finds the equivalent result that SDSS exhibits more
small scale power than 2dFGRS.
This behaviour is exactly what one expects if the more strongly
clustered red galaxies live in denser environments where the effects
of nonlinearity are greater.

This comparsion has revealed that to get unbiased estimates
of the cosmological parameters it is necessary to better
understand and constrain the distortion in the shape
of the power spectrum caused by nonlinearity and scale
dependent bias.  In \citet{cole05} a first attempt at modelling
this distortion was introduced using the $Q$ and $A$ parameters
of  equation~(\ref{eqn:nonlin}). For the mix of red and blue galaxies
present in the 2dFGRS survey, the necessary value of $Q$ was reasonably
small and the hence the scale dependent correction quite
modest. However for samples of redder or more luminous and hence
more clustered galaxies one expects greater nonlinearity and 
either greater values of $Q$\footnote{Subsequent to the presentation
of this talk on the 1st August 2007, three preprints analyzing
the SDSS-DR5 \citep{tegmark07,percival07a,percival07b} appeared
on astro-ph. As anticipated here, strong scale dependent bias was found
necessary to reconcile the measure power spectrum with linear theory
\citep[e.g $Q=26$ in][]{percival07b}}
 or the possible breakdown of this simple model.
Thus to get robust constraints from the main SDSS survey and in
particular the SDSS luminous red galaxy survey will require
more detailed modelling of nonlinearity and scale dependent bias.




\begin{thebibliography}{}

\bibitem[Cole et al.(2005)]{cole05}
Cole S., et al 2005, MNRAS, 362, 505     

\bibitem[Cross et al.(2004)]{cross}
Cross, N. J. G., Driver, S. P., Liske, J., Lemon, D. J.,
Peacock, J. A., Cole, S., Norberg, P., Sutherland, W. J., 2004, MNRAS, 349, 576

\bibitem[Eisenstein \& Hu(1998)]{EH98}
Eisenstein D.J., Hu, W., 1998, ApJ. 496, 605

\bibitem[Lewis \& Bridle(2002)]{lewis}
Lewis A., Bridle, S., 2002, Phys. Rev. D, 66, 103511   

\bibitem[Maddox et al.(1990)]{maddox90}
Maddox, S. J. Efstathiou, G., Sutherland, W. J., Loveday, J., 1990, MNRAS, 243, 692

\bibitem[Norberg et al.(2002)]{norberg02} 
Norberg P., et al., 2002, MNRAS, 336, 907

\bibitem[Percival, Verde \& Peacock(2004)]{PVP}
Percival W.J., Verde L., Peacock J.A., 2004, MNRAS, 347, 645

\bibitem[Percival et al.(2007a)]{percival07a} 
Percival W.J., et al., 2007a, astro-ph/0608635

\bibitem[Percival et al.(2007b)]{percival07b} 
Percival W.J., et al., 2007b, astro-ph/0608636

\bibitem[S\'anchez et al.(2006)]{sanchez06}
S\'anchez, A.G., et al 2006, MNRAS, 366, 189  

\bibitem[Stoughton et al.(2002)]{stoughton}
Stoughton, C. et al., 2002, AJ, 123, 485

\bibitem[Strauss et al.(2002)]{strauss02}
Strauss, M. et al., 2002, AJ, 124, 1810

\bibitem[Tegmark et al.(2004)]{tegmark04} 
Tegmark M., et al., 2004, ApJ, 606, 702

\bibitem[Tegmark et al.(2007)]{tegmark07} 
Tegmark M., et al., 2007, astro-ph/0608632








\end{thebibliography}
\end{document}